\documentclass[10pt]{iopart}
\usepackage{graphicx}
\begin{document}
\newcommand{\be}{\begin{equation}}
\newcommand{\ee}{\end{equation}}
\newcommand{\bea}{\begin{eqnarray}}
\newcommand{\eea}{\end{eqnarray}}
\title[Feasibility of detecting single atoms using photonic bandgap cavities]{Feasibility of detecting single atoms using photonic bandgap cavities}
\author{Benjamin Lev$^1$, Kartik Srinivasan$^2$, Paul Barclay$^2$, Oskar Painter$^2$ and Hideo Mabuchi$^1$}
\address{$^1$ Norman Bridge Laboratory of Physics, California Institute of Technology, Pasadena, CA 91125, USA}
\address{$^2$ Department of Applied Physics, California Institute of Technology, Pasadena, CA 91125, USA}
\ead{benlev@caltech.edu}
\begin{abstract}  We propose an atom-cavity chip that combines laser cooling and trapping of neutral atoms with magnetic microtraps and waveguides to deliver a cold atom to the mode of a fiber taper coupled photonic bandgap (PBG) cavity.  The feasibility of this device for detecting single atoms is analyzed using both a semi-classical treatment and an unconditional master equation approach.  Single-atom detection seems achievable in an initial experiment involving the non-deterministic delivery of weakly trapped atoms into the mode of the PBG cavity.
\end{abstract}
\pacs{42.50.Ct, 32.80.Pj, 03.75.Be, 42.70.Qs}

\section{Introduction}
The development of techniques necessary to manipulate single atoms and photons and to control their interactions is an important addition to the toolbox of nanotechnology.  An important advance would be the development of a compact and integrable device to serve as a single atom detector~\cite{Hinds03,Roman03}.  The system comprised of a strongly interacting atom and photon---cavity quantum electrodynamics (QED)~\cite{Kimble94,Kimble98,MabuchiSci}---provides the basis for realizing such a device.  These single atom detectors could play as important a role in the burgeoning field of atom optics~\cite{SpecialIssue} as single photon detectors do in conventional optics.  The advent of Bose-Einstein condensates (BECs) of neutral atoms and the production of degenerate fermionic condensates~\cite{Jin} further highlights the importance of developing single atom read-out devices.  

To achieve these goals in cavity QED, a neutral atom must be inside the mode of a high-finesse cavity with small mode volume:  the atom-cavity system must be in the strong coupling regime.  Strong coupling requires the atom-cavity coupling, $g_0$, to be much larger than both the atomic dipole decay rate, $\gamma_\perp$, and the decay rate of the cavity field, $\kappa$.  Specifically, the saturation photon number, $m_0=\gamma_\perp^2/2g_0^2$, and the critical atom number, $N_0=2\gamma_\perp \kappa/g_0^2$, must both be much less than unity.  

State-of-the-art cavity QED experiments have achieved strong-coupling parameters as small as $\left[m_0, N_0\right]\approx\left[10^{-4},10^{-3}\right]$ by either dropping~\cite{Hood00}, or vertically tossing~\cite{Rempe00} a cold neutral atom between the mirrors of a high-finesse, low-mode volume Fabry-Perot cavity.  Recently, intracavity atom trapping for durations up to 3 s has been demonstrated by coupling a secondary optical beam into the Fabry-Perot cavity to form a Far Off Resonance Trap (FORT)~\cite{McKeever03}.  

The intent of this paper is to introduce a cavity QED system based on magnetostatic delivery of atoms to a photonic bandgap cavity, and to discuss the ability of this system to detect single atoms.  This experimental system---magnetostatic confinement of atoms inside the field modes of photonic bandgap cavities---raises the possibility of achieving an experimentally robust, integrated, and scalable system.  Mastering the integration of a single atom and photons---quintessentially quantum components---presents an entirely new prospect for technology:  quantum computation and communication.  Cavity QED provides a rich experimental setting for quantum information processing (QIP), both in the implementation of quantum logic gates and in the development of quantum networks~\cite{Zoller95,Mabuchi01}.  While not necessary for single atom detection, confining the atom in the Lamb-Dicke regime inside the cavity for long periods of time is an important step towards accomplishing QIP using cavity QED.  An atom is trapped in the Lamb-Dicke regime when its recoil energy is less than the trap's vibrational level spacing, $\eta=(E_{recoil}/E_{vib})^{1/2}<1$.  This regime has been achieved using a FORT~\cite{McKeever03}, and magnetic microwire traps---such as those discussed in this paper---may also be capable of trapping atoms three-dimensionally inside a cavity in the Lamb-Dicke regime~\cite{Libbrecht,Mabuchi01}.

\section{Magnetic microtraps and photonic bandgap cavities}

Patterns of micron-sized wires can create magnetic field gradients and curvatures sufficiently large to accurately guide and trap atoms above the surface of the substrate~\cite{Libbrecht}.  These magnetic microtrap devices---commonly known as atom chips~\cite{Folman02,Jakob01a}---can be fabricated using standard photolithography techniques~\cite{Westervelt98,Lev03} and have been successfully used not only to trap and waveguide neutral atoms, but also to create and manipulate Bose-Einstein condensates~\cite{Jakob01b,Zimmermann01}.

Atom chips exploit the interaction potential, $V=-\vec{\mu}\cdot\vec{B}$, between an atom's magnetic moment, $\vec{\mu}$, and a wire's magnetic field, $\vec{B}$, to trap or guide weak-field seeking states of a neutral atom.  The simplest example of a magnetic microtrap involves the combination of the field from a U-shaped wire with a homogenous bias field, $B_{bias}$~\cite{Jakob99}.  The bias field, parallel to the wire substrate and perpendicular to the base of the U-wire, serves to cancel the curling field of the wire to form a two-dimensional quadrupole trap for the weak-field seeking atoms.  The atoms are confined in the third dimension by the fields from the side wires of the U-trap, forming a cigar-shaped trap above the wire surface.  The position of the trap minimum above the wire surface, $r$, and the gradient of the trap are completely determined by the magnitude of $B_{bias}$ and the current, $I$, in the U-wire,
\bea
r=\frac{\mu_0}{2\pi}\frac{I}{B_{bias}}, &\qquad \nabla B=\frac{2\pi}{\mu_0}\frac{B_{bias}^2}{I}.
\eea
For example, with a wire current of 1 A and a bias field of 10 G, the atoms are trapped 200 $\mu$m above the surface in a field gradient---perpendicular to the base of the U-wire---of 500 G/cm.  Ioffe traps---which are insusceptible to trap losses due to Majorana spin flips---may be formed either by a similar Z-trap~\cite{Jakob99} or by using wires forming patterns of nested arcs~\cite{Libbrecht}.  Although this latter Ioffe trap is more complicated, it does allow the possibility of trapping atoms three-dimensionally in the Lamb-Dicke regime inside a photonic bandgap cavity coplanar with the wires~\cite{Mabuchi01}. Simple waveguides for the atoms can be formed from the Z-trap by extending the base of the Z-wire, allowing the atoms to ballistically expand along the field minimum above the elongated wire.  Beam splitters and conveyor belts have been demonstrated using similar techniques~\cite{Folman02,Jakob01a}.

Standard laser cooling and trapping techniques~\cite{Metcalf} are used to load cold atoms into the magnetic microtraps and waveguides.  Typically, atoms are collected in a variant of the magneto-optical trap (MOT) that uses the atom chip surface as a mirror to form four of the six required laser cooling beams~\cite{Jakob99}.  This mirror MOT and subsequent sub-doppler cooling allows the collection of $10^6$, $\sim10$ $\mu$K, atoms a few millimeters above the chip's surface.  Conveniently, the quadrupole field from the U-trap is in the same orientation as the magnetic field required to form a mirror MOT.  In the most simple case, the atoms can be transfered to the U-trap by replacing the mirror MOT's quadrupole field with that of the U-trap while maintaining the cooling lasers in the same configuration:  this creates a U-MOT using the microwire magnetic field.  An alternative and more experimentally compact and robust method---and the one employed in our lab---traps the atoms directly from vapor using a large copper U-shaped block carrying 30 A and located underneath the atom chip~\cite{schmeid03}.  The atoms in this macro U-MOT are subsequently transfered to smaller, magnetostatic U-traps on the atom chip surface.  

The proximity of the atoms to the chip's surface naturally facilitates the integration of magnetically trapped atoms with on-chip cavities such as microdisks or photonic crystals.  Two-dimensional photonic bandgap (PBG) cavities---perforated semiconductor structures that confine light through the dual action of distributed Bragg reflection and internal reflection---are in many respects ideal for cavity QED~\cite{Jelena}.  Their small mode volume and modest quality factors open the possibility of achieving extremely small strong coupling parameters:  $\left[m_0,N_0\right]=\left[10^{-8},10^{-4}\right]$.  With regards to atom-cavity coupling, these cavities have the advantage over microdisks and microspheres in that the mode's field maximum can be located in the holes rather than inside the dielectric material.  As an inherently stable, monolithic structure, PBG cavities will not need the support structure for active stabilization that Fabry-Perot cavities require.  Moreover, their compactness and compatibility with fiber optics-based input and output couplers~\cite{barclay03,barclay03a} allows one to envision an array of PBG cavities, atom microtraps, input/output couplers, and other processing devices all on the same integrated chip. 

We plan to use PBG cavities of the graded defect design discussed in reference~\cite{kartik03}, which consist
of a rectangular lattice of air holes in an optically thin, high refractive index slab waveguide.Ê The holes 
gradually decrease in diameter towards the cavity center, and experimental measurements of such cavities
fabricated in silicon membranes (see figure 1(a)) and operating at $\lambda\sim1.6$ microns possess $Q$'s as 
high as 40,000 with modal volumes of $V_{eff}\sim0.9$ cubic wavelengths $(\lambda/n)^3$~\cite{kartik03a}.Ê In future experiments with single atoms, cavities will be etched in a thin AlGaAs membrane, chosen for its transparency at the 
wavelength of cesium's D2 transition, 852 nm.Ê For the $Q$ and $V_{eff}$ values mentioned above, the atom-cavity 
coupling can be a high as $g_0=2\pi\cdot17$ GHz while the decoherence rates are $[\kappa,\gamma_{perp}]/2\pi = [4.4\mbox{ GHz}, 2.6\mbox{ MHz}]$. This gives strong coupling parameters of $\left[m_0,N_0\right]=\left[1.2\times10^{-8},8.4\times10^{-5}\right]$, which are much smaller than those achieved in recent experiments using Fabry-Perot cavities, $\left[m_0,N_0\right]=\left[2.8\times10^{-4},6.1\times10^{-3}\right]$~\cite{Hood00}.  The central hole diameter is $\sim$100 nm and the membrane thickness is $\sim$170 nm.  An atom in this small hole will be affected by the Casimir-Polder potential~\cite{Vladan}, and cavity QED dynamics in the presence of this force will need to be investigated.

\begin{figure}
\begin{center}
\scalebox{.475}[.475]{\includegraphics{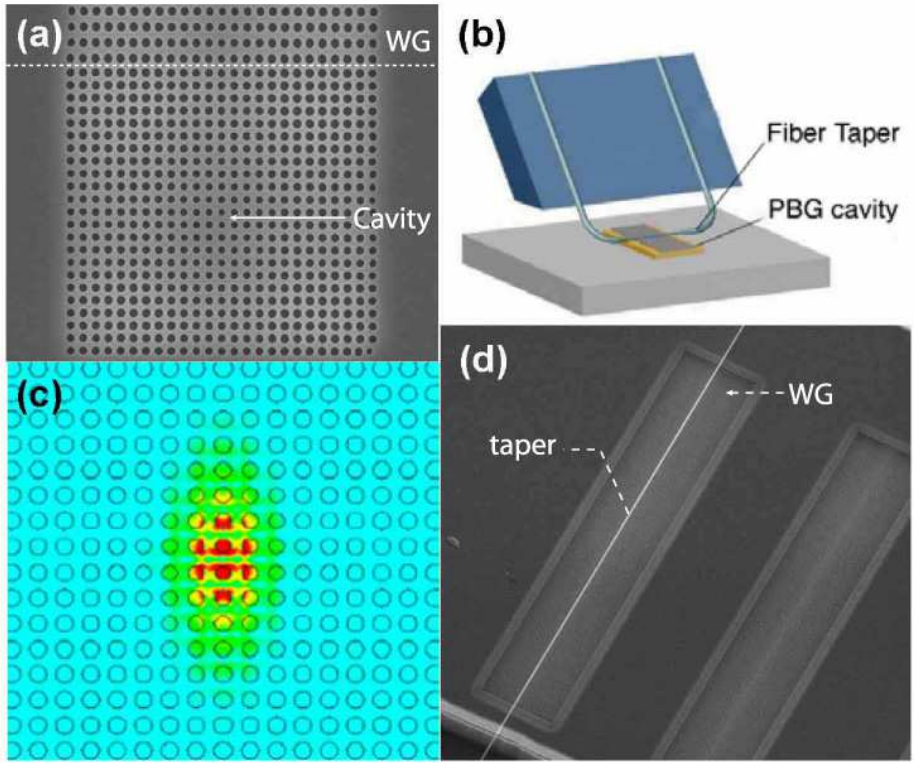}}
\end{center}
\caption{\label{fig:cavity} (a) Scanning electron microscope (SEM) image of a photonic bandgap cavity and waveguide (WG) fabricated in silicon.  (b) Schematic of the fiber taper coupler.  (c) Finite-difference time-domain calculated electric field amplitude of the cavity mode taken in the center of the membrane.  (d) SEM image of an optical fiber taper aligned above a photonic crystal waveguide.}
\end{figure}

The cavity is coupled to a photonic crystal waveguide, which in turn is evanescently coupled to an optical fiber taper.  By positioning the fiber taper---whose minimum diameter is on the order of a micron---in the near field of the photonic crystal waveguide and aligned along its axis (see figures~\ref{fig:cavity} (b) and (d)), highly efficient (greater than 98\%) fiber coupling into and out of the photonic crystal waveguide can be achieved~\cite{barclay03a}.  Light coupled into the photonic crystal waveguide is reflected by the PBG cavity and recollected in the backward propagating fiber taper mode~\cite{barclay03b}.  Figure~\ref{fig:cavity}(a) shows the boundary between the waveguide and the cavity:  the top four rows of holes are the end of the waveguide, which is formed in a similar fashion to the cavity, except that the holes are graded in only the lateral dimension.  This design maximizes the mode matching between the waveguide and the cavity modes~\cite{barclay03}.  The waveguide may be bent to allow access to the cavity unencumbered by the fiber.

\section{Experimental proposal}

\begin{figure}
\begin{center}
\scalebox{.75}[.75]{\includegraphics{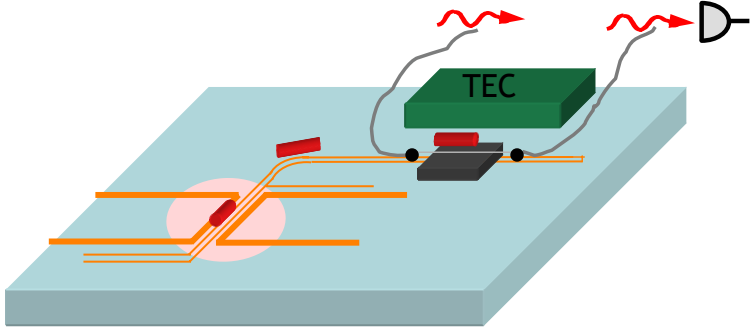}}
\end{center}
\caption{\label{fig:chip} Schematic of the atom-cavity chip experiment.  The microwire U-traps and atomic waveguides are shown as yellow wires, and the light red area centered about the U-traps represents the footprint of the reflected trapping lasers.  The atoms are the red cylinders, pictured as they are transported towards the PBG cavity which is shown as the black chip glued to the substrate's surface.  The grey line is the optical fiber and fiber taper.}
\end{figure}

As a first generation experiment, we would like to bring a trapped cloud of cold neutral atoms---cesium in our case---into contact with a PBG cavity, simultaneously demonstrating the integration of a cavity with an atom chip and the strong coupling of a neutral atom to a PBG cavity.  Figure~\ref{fig:chip} shows a rough schematic of the atom-cavity chip experiment.  
The chip is divided into two regions, one for laser trapping and cooling of the atoms in a U-MOT and U-traps, and the other for the PBG cavity and its tapered fiber and photonic crystal waveguide couplers.  The two regions are connected by a microwire waveguide to transport the atoms from the laser cooling region to the PBG cavity. These regions must be separated by 1 to 2 cm in order for the bulk of the cavity to not obstruct the 1 cm$^2$ U-MOT beams.  Furthermore, to position the cavity outside of the horizontal U-MOT beam that grazes the substrate surface, the waveguide must convey the atoms around a $90^\circ$ turn.  This will be accomplished either by using a two-wire guide (chosen for depiction in figure~\ref{fig:chip} for simplicity of illustration)~\cite{schmiedspiral} or by rotating the atoms in a P-trap---similar to a U-trap but with the base wire bent allowing a rotating bias field to change the orientation of the atoms~\cite{JakobPrivate}---before transferring the atoms into a Z-trap waveguide aligned perpendicular to the initial U-trap.  This latter design has the advantage that the simple addition of a few coplanar wires can serve to loosely confine the atoms once they reach the PBG cavity.

In the PBG region, the atoms are suspended a few 100 microns above the surface of the waveguide's microwires, and this allows enough room for the $\sim$200 micron thin PBG substrate to be placed in the gap between the atoms and the microwires.  Once the atoms are transported to a position above the PBG cavity, the current and bias field of the guide are adjusted to lower the cold atom cloud into the surface of the PBG cavity.  A thermoelectric cooler (TEC) is located near the PBG cavity to counteract heating due to the microwire waveguide, maintaining a specific cavity detuning from the frequency driving laser and the atomic resonance.  We estimate a cavity tunability of 20 GHz/$^\circ$C, and with TEC control of 10$^{-2}$ $^\circ$C, we should be able to achieve a 200 MHz tuning resolution.  This resolution is sufficient, as we expect to operate with detunings on the order of 1 to 10 GHz. 


The delivery scheme described above provides a non-deterministic source of weakly trapped atoms to the cavity mode.  The field of the cavity mode is concentrated in the central $\sim$10 holes (see Figure~\ref{fig:cavity} (c)).  We expect to transport 10$^5$ atoms in a cigar-shaped cloud of density ~10$^{11}$/cm$^3$.  The cross-sectional area of this cloud parallel to the chip is larger than the 0.4 $\mu$m$^2$ area of the PBG cavity that is occupied by the field, and we estimate there is a $\sim$10\% probability of an atom encountering one of the central 10 holes per cloud interaction.  With an experimental repetition once every $\sim$5 seconds---limited by the U-MOT replenishing time---we foresee the accumulation of a significant number of events in a reasonable amount of time, and as discussed in section~\ref{OBE} below, we expect to detect strong signals during single atom transits through the PBG cavity's central holes.  If we assume a cesium cloud temperature of 10 $\mu$K, then a cesium atom whose velocity is parallel to the axis, $\hat{z}$, of a central hole will interact with the mode for a time duration of $\sim10$ $\mu$s. 

\section{Single atom detectability}\label{OBE}

To investigate the PBG cavity's response to a strongly coupled atom falling through a central hole, we solve---using a two-level atom---the semiclassical optical bistability equation for a qualitative understanding of the interaction and the quantum master equation to obtain a more quantitative description.  Although neither of these treatments fully encompasses the complexity of the system, we presume that they are sufficient for demonstrating the feasibility of the device for single atom detection.  These calculations ignore the fact that $g_0$ and the detunings are of the same order or much larger than both the hyperfine ground-state and excited-state splittings, which for cesium are 9.2 GHz and 151 to 251 MHz, respectively.  In other words, the atom-photon coupling is much stronger than the coupling between the electron and nuclear spins.  This is an unusual situation and requires a full quantum calculation of the atom-PBG cavity interaction that includes the full cesium D2 manifold of states.  We are in the process of performing this computation.

\begin{figure}[h]
\begin{center}
\scalebox{.5}[.5]{\includegraphics{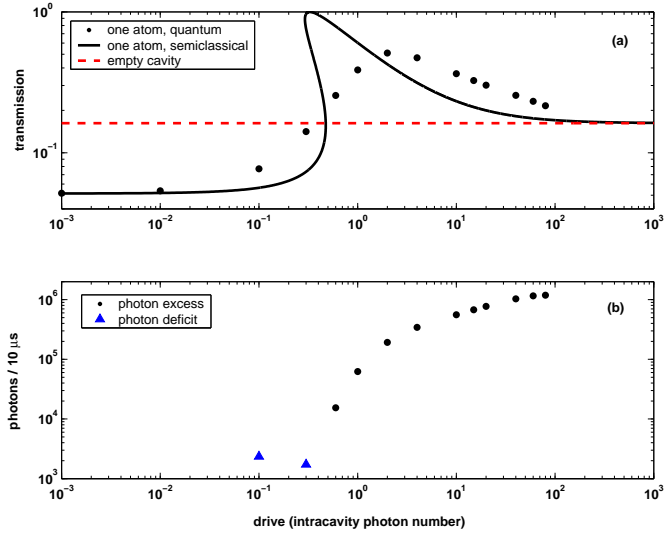}}
\end{center}
\caption{\label{fig:10a10c} (a) Transmisson of the cavity as a function of drive strength---measured in intracavity photon number for a resonant and empty cavity---calculated from equation~\ref{OBeqn} (black line) and from equation~\ref{ME} (points).  The empty cavity transmission is shown as a dashed red line.  (b) Difference in output---during the expected 10 $\mu$s of atom-cavity interaction---between a cavity with one atom and an empty cavity.  The detunings are $[\Delta,\theta]/2\pi=[10,10]$ GHz.}
\end{figure}

\begin{figure}[h]
\begin{center}
\scalebox{.5}[.5]{\includegraphics{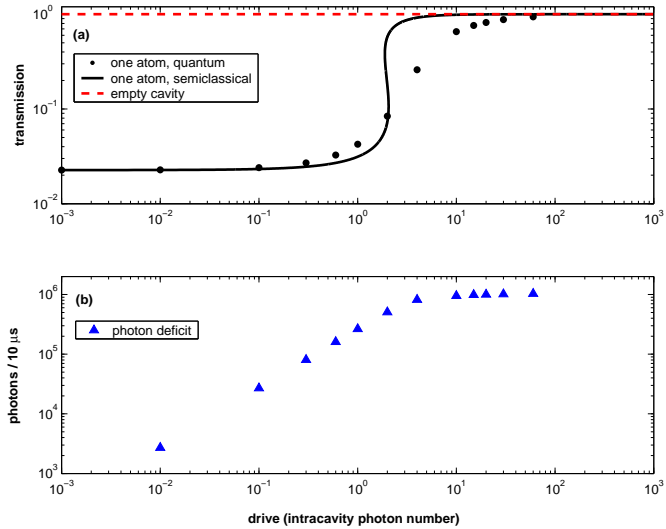}}
\end{center}
\caption{\label{fig:10a0c} (a) and (b) same as figures~\ref{fig:10a10c} (a) and (b) except with detunings of $[\Delta,\theta]/2\pi=[10,0]$ GHz.}
\end{figure}

The optical bistability equation is a semiclassical description of the transmission of a cavity containing atoms~\cite{Lugiato}, 
\be \label{OBeqn}
y=\frac{x}{\left[\left(1+\frac{2}{N_0(1+(\Delta/\gamma)^2+y^2)}\right)^2+i\left(\frac{\theta}{\kappa}-\frac{2\Delta}{\gamma N_0(1+(\Delta/\gamma)^2+y^2)}\right)^2\right]^{\frac{1}{2}}}.
\ee
In the above equation, $x$ is the input field, $E/\sqrt{m_0}$, where $E$ is the amplitude of the driving field;  $y$ is the output field, $\alpha/\sqrt{m_0}$, where $\alpha$ is the intracavity coherent state amplitude; $\Delta$ is the atom-laser detuning; and $\theta$ is the cavity-laser detuning.  The black curves in figures~\ref{fig:10a10c} (a) and~\ref{fig:10a0c} (a) show the solution to equation~\ref{OBeqn} for $[\Delta,\theta]/2\pi=[10,10]$ GHz and $[\Delta,\theta]/2\pi=[10,0]$ GHz, respectively.  These two sets of detunings are chosen to highlight different atom-cavity response regimes where we expect to be able to detect single atoms.  In both plots, $[g_0,\kappa,\gamma_{perp}]/2\pi = [17\mbox{ GHz}, 4.4\mbox{ GHz}, 2.6\mbox{ MHz}]$.  The horizontal dashed lines are the empty cavity transmissions.  Both semiclassical solutions show signs of bistability in the region around one intracavity photon.  Within the context of the approximation of equation~\ref{OBeqn}, figure~\ref{fig:10a10c} (a) shows that for 10 GHz detunings of the atom and cavity from the probe laser, an excess of photons transmitted through the cavity---an ``up-transit"---can be detected for a drive of a few intracavity photons.  Figure~\ref{fig:10a0c} (a) shows that with the cavity on resonance with the laser and the atom 10 GHz detuned, a deficit of photons---a ``down-transit"--- can be detected for similar drive strengths of a few intracavity photons.

\begin{figure}
\begin{center}
\scalebox{.5}[.5]{\includegraphics{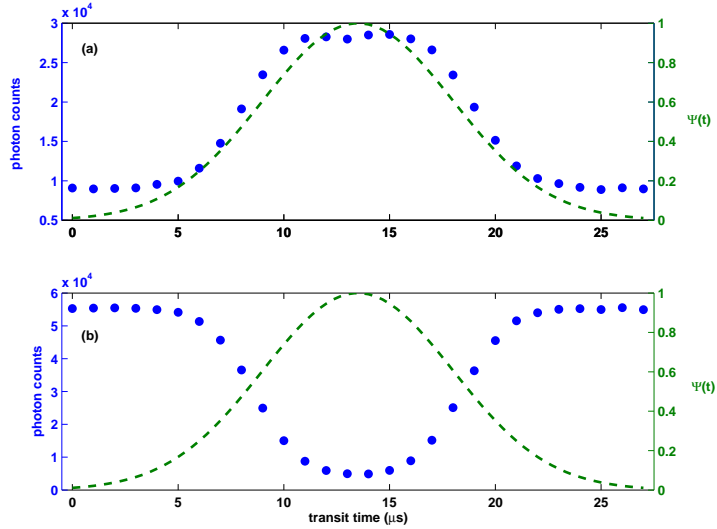}}
\end{center}
\caption{\label{fig:transits} Simulated photon counts due to atoms transits through the axis of the cavity's central hole.  Blue dots (left axis) are the photon counts, and the green, dashed curve (right axis) is the gaussian variation of $g(t)/g_0=\psi(z(t))$ experienced by the atom during its transit.  Calculations are for detunings of (a) $[\Delta,\theta]/2\pi=[10,10]$ GHz and (b) $[\Delta,\theta]/2\pi=[10,0]$ GHz.}
\end{figure}

The solutions to the unconditional master equation paint a more accurate picture of the atom-cavity system.  Under the two-level atom, electric dipole, and rotating-wave approximations, the equation for the density matrix, $\rho$, of the joint state of the atom and cavity is as follows: 
\bea
\dot{\rho}&=&\frac{-i}{\hbar}\left[\hat{H}_0,\rho\right]+\gamma_{\perp}(2\hat{\sigma}\rho\hat{\sigma}^\dagger-\hat{\sigma}^\dagger\hat{\sigma}\rho-\rho\hat{\sigma}^\dagger\hat{\sigma}) \nonumber \\& &+\kappa(2\hat{a}\rho\hat{a}^\dagger-\hat{a}^\dagger\hat{a}\rho-\rho\hat{a}^\dagger\hat{a}), \label{ME}\\
\hat{H}_0&=&\hbar\Delta\hat{\sigma}^\dagger\hat{\sigma}+\hbar\theta\hat{a}^\dagger\hat{a}+i\hbar E(\hat{a}^\dagger-\hat{a})+\hat{H}_{int}, \\
\hat{H}_{int}&=& i\hbar g_0\psi(\hat{r})\left[\hat{a}^\dagger\hat{\sigma}-\hat{\sigma}^\dagger\hat{a}\right].
\eea
In this equation, $\hat{\sigma}$ is the atomic lowering operator and $\hat{a}$ is the cavity field annihilation operator.  Along the axis of the central cavity hole, the mode function, $\psi(z)$, closely approximates a gaussian of width $\sim$225 nm, centered about the midpoint of the $\sim$170 nm thick cavity membrane.  The steady-state density operator, $\rho_{ss}$, as a function of various drive strengths, coupling strengths, and detunings is found by solving equation~\ref{ME} with $\dot{\rho}_{ss}=0$.  Operator expectations are $\langle\hat{O}\rangle=\mbox{Tr}[\rho_{ss}\hat{O}]$.  The expected cavity output in photons per detector integration time, $\Delta t$, is 
\be
N=\kappa \Delta t \langle\hat{a}^\dagger\hat{a}\rangle,
\ee
with noise fluctuations of variance
\be
(\Delta N)^2=\kappa \Delta t (\langle\hat{a}^\dagger\hat{a}\hat{a}^\dagger\hat{a}\rangle-\langle\hat{a}^\dagger\hat{a}\rangle^2).
\ee
Note that instead of photon counting, heterodyne detection may be used, in which case expectations of $\hat{a}$ rather than $\hat{a}^\dagger\hat{a}$ are the relevant quantities.  The results presented in figures~\ref{fig:10a10c} through~\ref{fig:transits} are qualitatively similar for either case.

The points in figures~\ref{fig:10a10c} (a) and~\ref{fig:10a0c} (a) represent $N$ calculated from solutions to equation~\ref{ME} for various drive strengths and for the same $g_0$, $\kappa$, and detunings as used in the semiclassical calculation.  These points do not extend past a drive strength of 80 intracavity photons because our limited computational resources necessitate the use of a truncated Fock basis.  The cavity transmission as a function of drive qualitatively follows the semiclassical solutions, however, there is no longer a sign of bistability, which is to be expected since the unconditional master equation is linear in the state variables, $\rho$, and we plot only $\langle N \rangle$.  We also see that for a drive of 1 to 10 photons, up-transits occur for a probe laser detuned 10 GHz from both the atom and the cavity (figure~\ref{fig:10a10c} (a)), and down-transits for a probe laser and cavity 10 GHz detuned from the atom (figure~\ref{fig:10a0c} (a)).  Figures~\ref{fig:10a10c} (b) and~\ref{fig:10a0c} (b) show the change in the output of the cavity---using the master equation solutions---during the 10 $\mu$s we expect the atom to interact with the cavity mode.  The black dots show the up-transits and blue triangles the down-transits.  For drive powers of $\sim1$ nW (1 to 10 intracavity photons), photon excesses of $10^5$ to $10^6$ can be seen in the up-transits of the $[\Delta,\theta]/2\pi=[10,10]$ GHz case (figure~\ref{fig:10a10c} (b)), and photon deficits of $10^5$ to $10^6$ in the down-transits of the $[\Delta,\theta]/2\pi=[10,0]$ GHz case (figure~\ref{fig:10a0c} (b)).  For both sets of detunings, we see that for drive strengths less (greater) than one intracavity photon, there are super- (sub-) Poissonian noise fluctuations of the photon number.  Plots of the Q-function~\cite{M&W} in the sub-Poissonian regions show excess spread---and even a bifurcation in the $[\Delta,\theta]/2\pi=[10,0]$ GHz case---of the phase quadrature corresponding to photon number squeezing.

Simulated photon counts during atoms transits are shown in figures~\ref{fig:transits} (a) and (b).  We assume the atom moves with constant velocity, $v=2.5$ cm/s, through the axis of the cavity mode $\psi(z)$, making a full transit of the gaussian waist in 10 $\mu$s.  In both plots the drive strength is 2 intracavity photons.  As the atom transverses the cavity, the coupling $g(t)=g_0\psi(vt)$ also varies as a gaussian, which modulates the output photon flux.  The mean photon count, $N$, and variance, $(\Delta N)^2$, are found by solving for $\rho_{ss}$ for each $g(t)$ in time steps of $\Delta t=1$ $\mu$s, chosen to simulate a finite bandwidth photodetector.  Each point includes additional shot-noise selected randomly from a normal distribution of standard deviation $\Delta N$.  Figures~\ref{fig:transits} (a) and (b) show that even with shot-noise, up- and down-transits of single atoms through the axis of the central PGB cavity hole are clearly detectable.  Moreover, it seems possible to detect atom transits that only experience 20\% to 30\% of $g_0$.  During an experiment, we expect to detect a low background of signals from marginally coupled atoms---such as those grazing the field extending from the surface of the PBG membrane or slipping into holes away from the central region---punctuated by sharp pikes representing atoms fully coupled to the field inside the central holes.  It should be noted that the mean photon numbers and noise in figures ~\ref{fig:transits} (a) and (b) are not derived from a quantum trajectory calculated from the conditional master equation~\cite{carmichael}, but are simply calculated using $\rho_{ss}$ from the unconditional equation~\ref{ME}.  This is acceptable given the inherent limitations of the model as mentioned at the beginning of this section.

The atom will experience a force,
\be
\langle \vec{f} \rangle=-i\hbar\nabla g(\vec{r})\langle  \hat{a}^\dagger\hat{\sigma}- \hat{a}\hat{\sigma}^\dagger\rangle,
\ee
as it encounters the cavity mode.  The maximum acceleration on an atom dragged though the cavity mode at velocity 2.5 cm/s---for either of the sets of detunings used above---is $|\langle f_{max} \rangle |/M_{Cs}=2.4\times10^8$ m/s$^2$, corresponding to a change in velocity of 
\be
\Delta v =\sqrt{\frac{|\langle f_{max} \rangle |\Delta z}{M_{Cs}}}\approx5\mbox{ m/s}
\ee
over half the length of the cavity mode, $\Delta z = 100$ nm.  In the above equations, $M_{Cs}$ is the mass of a cesium atom.  This agrees with a simple estimate using 
\be \label{simple}
\hbar g_0=0.5M_{Cs}(\Delta v)^2,
\ee
which yields $\Delta v = 10$ m/s.  Fabry-Perot experiments have detected effects of the cavity interaction on the atomic motion~\cite{mabuchiOL}.  The simple estimate using equation~\ref{simple} gives a smaller value of $\Delta v \approx0.7$ m/s for the Fabry-Perot experiments, implying that the motion of the atom traversing the mode of the PBG cavity will also be significantly affected.  A more detailed calculation~\cite{dodog} of the force and momentum diffusion using a master equation beyond the two-level atom approximation is necessary to make predictions about the behavior of an atom in an attractive, red-detuned cavity mode or in a repulsive, blue-detuned mode.  The close proximity of the atom to the sides of the PGB cavity's holes will surely affect the system's dynamics due to the Casimir-Polder potential~\cite{Vladan}, and this will need to be addressed in more detailed simulations.
 
\section{Conclusion}

The integration of atom trapping and cooling with photonic bandgap cavities on a chip introduces a robust and scalable cavity QED system to the toolbox of nanotechnology.   A device allowing cooled neutral atoms to be delivered via a magnetic microtrap and waveguide to the mode of a graded lattice PBG cavity is feasible given present technology.  Calculations using the semi-classical optical bistability equation and the unconditional master equation indicate that it will be possible to detect single strongly-coupled atoms with this atom-cavity chip.  

\ack
This work was supported by the Multidisciplinary University Research Initiative program under Grant No. DAAD19-00-1-0374 and the Charles Lee Powell Foundation.  K.S. thanks the Hertz Foundation for financial support.


\section*{References}
\begin{thebibliography}{99}

\bibitem{Hinds03}  Horak P, Klappauf B G, Haase A, Folman R, Schmiedmayer J, Domokos P and Hinds, E A 2003 {\it \PR A} {\bf{67}} 043806

\bibitem{Roman03} Long R, Steinmetz T, Hommelhoff P, H\"{a}nsel W, H\"{a}nsch T W and Reichel J 2003 {\it Phil. Trans. R. Soc. Lond. A} {\bf{361}} 1

\bibitem{Kimble94} Kimble H J 1994 {\it Cavity Quantum Electrodynamics} (San Diego:  Academic Press, Edited by Berman P) p 203

\bibitem{Kimble98} Kimble H J 1998 {\it Physica Scripta} {\bf{T76}} 127

\bibitem{MabuchiSci} Mabuchi H and Doherty A C 2002 {\it Science} {\bf{298}} 1372

\bibitem{SpecialIssue} {\it Quantum Semiclass. Opt.} 1996 {\bf{8}}(3)

\bibitem{Jin} Regal C A, Greiner M and Jin D S 2004 \PRL {\bf{92}} 040403

\bibitem{Hood00} Hood C J, Lynn T W, Doherty A C, Parkins A S and Kimble H J 2000 {\it Science} {\bf{287}} 1447 

\bibitem{Rempe00} Pinkse P W H, Fisher T, Maunz P and Rempe G 2000 {\it Nature} {\bf{404}} 365

\bibitem{McKeever03} McKeever J, Buck J R, Boozer A D, Kuzmich A, N\"{a}gerl H C, Stamper-Kurn D M and Kimble H J 2003 {\it\PRL} {\bf{90}}
133602

\bibitem{Zoller95} Pellizzari T, Gardiner S A, Cirac J I and
Zoller P 1995 \PRL {\bf{75}} 3788

\bibitem{Mabuchi01} Mabuchi H, Armen M, Lev B, Loncar M, Vuckovic J, Kimble H J, Preskill J, Roukes M and Scherer A 2001 {\it Quantum Inf. Comput.}
{\bf{1}} 7

\bibitem{Libbrecht} Weinstein J D and Libbrecht K G 1995 {\it \PR A} {\bf{52}}
4004

\bibitem{Folman02} Folman R, Kr\"{u}ger P, Schmiedmayer J, Denschlag J and Henkel C
 2002 {\it Adv. At. Mol. Opt. Phys.} {\bf{48}} 263
 
\bibitem{Jakob01a} Reichel J, H\"{a}nsel W, Hommelhoff P and H\"{a}nsch T W 2001 {\it Appl. Phys. B} {\bf{72}} 81

\bibitem{Westervelt98} Johnson K S, Drndic M, Thywissen J H, Zabow G, Westervelt R M and Prentiss M 1998
{\it\PRL} {\bf{81}} 1137

\bibitem{Lev03} Lev B 2003 {\it Quantum Inf. Comput.} {\bf{3}} 450

\bibitem{Jakob01b} Hansel W, Hommelhoff P, H\"{a}nsch T W and Reichel J 2001 {\it Nature} {\bf{413}}
498

\bibitem{Zimmermann01} Ott H, Fortagh J, Schlotterbeck G, Grossmann A and Zimmermann C
2001 {\it\PRL} {\bf{87}} 230401

\bibitem{Jakob99} Reichel J, H\"{a}nsel W and H\"{a}nsch T W 1999 {\it\PRL} {\bf{83}}
3398

\bibitem{Metcalf} Metcalf H and van der Straten P 1999 {\it Laser Cooling and
Trapping} (New York: Springer-Verlag)

\bibitem{schmeid03} Wildermuth S, Kr\"{u}ger P, Becker C, Brajdic M, Haupt S, Kasper A, Folman R and Schmiedmayer J 2003 {\it Preprint} cond-mat/0311475

\bibitem{Jelena} Vu\^{c}kovi\'{c} J, Lon\^{c}ar M, Mabuchi H and Scherer A 2001 {\it\PR E} {\bf{65}} 016608

\bibitem{barclay03} Barclay P E, Srinivasan K, and Painter O 2003 {\it J. Opt. Soc. Am. B} {\bf{20}} 2274

\bibitem{barclay03a} Barclay P E, Srinivasan K, Borselli M and Painter O 2003 {\it Preprint} quant-ph/0308070

\bibitem{kartik03} Srinivasan K and Painter 0 2003 {\it Optics Express} {\bf{10}} 670

\bibitem{kartik03a} Srinivasan K, Barclay P E, Borselli M and Painter O 2003 {\it Preprint} quant-ph/0309190

\bibitem{barclay03b} Barclay P E, Srinivasan K, Borselli M and Painter O 2003 {\it Preprint} physics/0311006

\bibitem{Vladan} Lin Y, Teper I, Chin C and Vuleti\'{c} V 2003 {\it\PRL} {\bf{92}}
050404

\bibitem{schmiedspiral} Luo X, Kr\"{u}ger P, Brugger K, Wildermuth S, Gimpel H, Klein M W, Groth S, Folman R, Bar-Joseph I and Schmiedmayer J 2003 {\it Preprint} quant-ph/0311174

\bibitem{JakobPrivate} Long R and Reichel J 2002 Private communication

\bibitem{Lugiato} Lugiato L A 1984 {\it Progress in Optics} (Amsterdam:  Elsevier Science Publishers) p 71

\bibitem{M&W} Mandel L and Wolf E 1997 {\it Optical Coherence and Quantum Optics} (Cambridge:  Cambridge University Press)

\bibitem{carmichael} Carmichael H 1993 {\it An Open Systems Approach to Quantum Optics} (Berlin:  Springer-Verlag)

\bibitem{mabuchiOL} Mabuchi H, Ye J and Kimble H J 1999 {\it Appl. Phys. B} {\bf{68}} 1095

\bibitem{dodog} Doherty A C, Parkins A S, Tan S M and Walls D F 1996 {\it\PR A} {\bf{56}} 833

\end{thebibliography}
\end{document}